\documentclass[runningheads,openany]{svmult}

\usepackage{graphicx}  % standard LaTeX graphics tool
\usepackage{subeqnar}  % subnumbers individual equations
\usepackage{multicol}  % used for the two-column index
\usepackage{physprbb}  % modified textarea for proceedings,
\makeindex             % used for the subject index
\usepackage{epsfig}
\begin{document}

%%%%%%%%%%%%%%%%%%%%%%%%%%%%%%%%%%%%%%%%%%%%%%%%%%%%%%%%%%%%%

\renewcommand{\contriblistname}{List of Participants Contributed to this Edition}

\setlength{\floatsep}{24pt plus 4pt minus 4pt}

\title*{2s Hyperfine Structure\protect\newline 
in Hydrogen Atom and~Helium-3 Ion}
\toctitle{2s Hyperfine Structure in Hydrogen Atom and~Helium-3 Ion}
\titlerunning{2s Hyperfine Structure in Hydrogen and Helium}

\author{Savely G. Karshenboim\inst{1,2}\thanks{E-mail: sek@mpq.mpg.de}
}

\authorrunning{Savely G. Karshenboim}

\institute{
D. I. Mendeleev Institute for Metrology, 
198005 St. Petersburg, Russia
\and Max-Planck-Institut f\"ur Quantenoptik, 
85748 Garching, Germany
}

\maketitle

\begin{abstract}
The usefulness of study of hyperfine splitting in the hydrogen atom is limited on a level of 10 ppm
by our knowledge of the proton structure. One way to go beyond 10 ppm is to study a specific
difference of the hyperfine structure intervals $8 \Delta \nu_2 - \Delta \nu_1$. Nuclear effects for
are not important this difference and it is of use to study higher-order QED corrections.
\end{abstract}

\label{c_kar1}

\index{State-dependent corrections|(}
\index{Hydrogen atom!hyperfine structure|(}

\section{Introduction}

The hyperfine splitting of the ground state of the hydrogen atom has been for a while 
one of the most precisely known physical quantities,
\index{Bound state QED!precision tests}
however, its use for tests of QED theory is limited by a lack of our knowledge of the proton structure. 
The theoretical uncertainty due to that is on a level of 10 ppm. To go farther with theory we need to 
eliminate the influence of the nucleus. \index{Proton structure} A few ways have been used 
(see e. g. \cite{17icap}):
\begin{itemize}
\item to remove the proton from the hydrogen atom and to study a two-body system, 
which is like hydrogen, but without any nuclear structure, namely: muonium 
\cite{17jungmann} or positronium \cite{17conti}; 
\index{Muonium!hyperfine structure}\index{Positronium!hyperfine structure}%
\item to compare the hyperfine structure intervals of the $1s$ and $2s$ states (this work);
\index{Metastable levels!$2s$|(}
\item to measure the hyperfine splitting in muonic hydrogen and to compare it with the one in 
a normal hydrogen atom (a status report on the $n=2$ muonic hydrogen project is presented in Ref. 
\cite{17pohl}; comparison of the $1s$ and $2s$ hfs and possibility to measure $1s$ hfs 
is considered in Ref. \cite{17jik}.
\index{Muonic hydrogen!hyperfine structure}
\end{itemize}
Recently there has been considerable progress in 
measurement and calculation of
the hyperfine splitting of the ground state and  the $2s_{1/2}$
state in the hydrogen atom. The $2s_{1/2}$ 
hyperfine splitting in hydrogen was determined to be \cite{17rothery}
\begin{equation}
\Delta \nu_2({\rm H}) = 177\;556.785(29) ~{\rm kHz}\;,
\end{equation}
While less accurate than the classic determination of the ground state hyperfine
splitting, the combination of 1s and 2s {\em hfs} intervals
\begin{equation}
D_{21} ({\rm H}) = 8 \Delta \nu_2 - \Delta \nu_1\;.
\end{equation}
which is
determined in the hydrogen atom \cite{17rothery} as 
\[
D_{21}(H) = 48.528(232)~{\rm kHz}\;,
\]
has more implications for tests of bound state QED
because there is significantly less dependence on the poorly understood proton
structure contributions. Specifically, the theoretical uncertainty for the ground
state from the proton structure is about 10 kHz, while the
uncertainty for the combination is estimated to be few Hz.

\index{Helium ion!hyperfine structure in $^3$He$^+$|(}
On the theoretical front, there has been considerable progress in the calculation
of the ground state hyperfine splitting. Taken together 
with earlier
calculations of $D_{21}$ \cite{17zwanziger,17sternheim}, which were possible
because of cancellations of a number of large terms,
one can now give quite accurate values for $\Delta \nu_1$ and $\Delta \nu_2$,
We collect in Tables 1 and 2 along with the hydrogen results, the known experimental and
theoretical results for the deuterium atom and the $^3$He$^+$ ion. \index{Deuterium atom}
The helium results \cite{17he1s}
\begin{equation}
\Delta \nu_1(^3{\rm He}^+) = 8665\;649.867(10)~ {\rm kHz}
\end{equation} 
and \cite{17prior}
\begin{equation}
\Delta \nu_2(^3{\rm He}^+) = 1083\;354.969(30) ~{\rm kHz}
\end{equation}
lead to the most accurate value for the difference
\begin{equation}
D_{21}(^3{\rm He}^+) = 1\,189.979(71)~{\rm kHz}\;.
\end{equation}
\begin{table}
\caption{Comparison of the QED part of the theory to the experiment for hydrogen and 
deuterium atoms and for the $^3$He$^+$ ion. The results are presented in kHz}
\label{17Tab2}
\begin{center}
\def\arraystretch{1.4}
\setlength\tabcolsep{5pt}
\begin{tabular}{ccc@{\hskip 3em}cc}
\hline
Atom & \multicolumn{2}{c}{Experiment} & \multicolumn{2}{c}{QED theory for $D_{21}$} \\

     & $D_{21}({\rm exp})$ &  Refs.: 2s/1s             & Old & New \\
\hline
H         &  48.528(232)      & \protect{\cite{17rothery}}/\protect{\cite{17h1s}} & 48.943 &48.969(2)   \\
H         &    49.13(40)      & \protect{\cite{17h2s}}/\protect{\cite{17h1s}} &  &   \\
D         &  11.16(16)        & \protect{\cite{17d2s}}/\protect{\cite{17d1s}} & 11.307 & 11.3132(4)   \\
$^3$He$^+$  &  1\,189.979(71) & \protect{\cite{17prior}}/\protect{\cite{17he1s}} & 1\,189.795 & 1\,191.126(40) \\
$^3$He$^+$  &  1\,190.1(16)   & \protect{\cite{17he2s}}/\protect{\cite{17he1s}} &  &  \\
\hline
\end{tabular}
\end{center}
\end{table}

\index{Metastable levels!$2s$|)}

\section{Theory}

We consider a hydrogen-like system with 
a nucleus of charge $Z$, mass $M$, spin $I$, and magnetic moment $\mu$. 
The basic scale of the hyperfine splitting is then given by the Fermi formula,
\begin{equation}
E_F = {8 \over 3} \,Z^3 \alpha^2 \,Ryd \, { |\mu| \over \mu_B} \,{ 2 I + 1 \over 2 I}\,
\left({ M\over m+M}\right)^3\;.
\end{equation}
Here we take the fine structure constant $\alpha$ derived from g-2 value of electron  
$\alpha^{-1} = 137.035\;999\;58(52)$.
In addition we use
a value the Rydberg constant of 
$ Ryd = c \,Ry = 3.289\;841\;931 \cdot 10^{12}$ kHz.

We present the hyperfine structure as a sum
\begin{equation}
\Delta\nu_n = \Delta\nu_n(QED)+\Delta\nu_n({\rm nuclear~structure})\;.
\end{equation}

\subsection{Non-recoil limit}

\index{Relativistic corrections}
First we consider the external-field approximation. For a point-like
nucleus, they can be compactly parameterized by the equation
\[
\Delta \nu_n(\mbox{\rm N-R}) = {E_F \over n^3} \,\left[ B_n + {\alpha \over \pi} D^{(2)}_n(Z\alpha) +
\left({\alpha \over \pi}\right)^2 D^{(4)}_n(Z\alpha)+ ...\right]\;.
\]
Here, with  $\gamma = \sqrt{1-(Z\alpha)^2}$, \cite{17brei}
the Breit relativistic contribution is 
\begin{equation}
B_{1} = { 1 \over \gamma \,(2 \gamma -1)} \simeq 1 + {3 \over 2} \,(Z\alpha)^2 + {17 \over 8}\, 
(Z\alpha)^4 +...
\end{equation}
and
\begin{equation}
B_{2} = {2 \big(2(1+\gamma)+\sqrt{2(1+\gamma)}\big) \over 
(1+\gamma)^2\,\gamma \,(4 \gamma^2-1)} 
\simeq 1 + {17 \over 8}\, (Z\alpha)^2 + {449\over 128}\,(Z\alpha)^4 + ...
\end{equation}
and the functions
$D^{2r}_n(Z\alpha)$ represent $r$ loop radiative corrections. In the limit $Z \alpha=0$ they
reduce to the power series expansion of the electron $g\!-\!2$ factor, and the difference
is refered to a binding correction. For the ground state,
\index{Radiative corrections!01@one-loop}
\begin{eqnarray}
D^{(2)}_{1} &=&  {1 \over 2} + \pi (Z\alpha)\,\left(\ln(2) - { 5 \over 2}\right)   
+  (Z \alpha)^2\Biggl[ - {8 \over 3} {\rm ln}^2 (Z\alpha) \nonumber\\
&+& \left( {16 \over 3}\,  \ln(2) - 
{ 281 \over 180}\right){\rm ln}(Z \alpha) 
 + G^{\rm SE}_{1} (Z\alpha) + G^{\rm VP}_{1} (Z\alpha) \Biggr] 
\end{eqnarray}
and for the excited state
\begin{eqnarray}
D^{(2)}_{2} &=&  {1 \over 2} + \pi (Z\alpha)\,\left(\ln(2) - { 5 \over 2}\right) 
+  (Z \alpha)^2 \Biggl[ - {8 \over 3} {\rm ln}^2 (Z\alpha) 
\nonumber\\
&+& \left( {32 \over 3}\, \ln(2) - 
{ 1541 \over 180}\right){\rm ln}(Z \alpha) 
 + G^{\rm SE}_{2} (Z\alpha) + G^{\rm VP}_{2} (Z\alpha) \Biggr] \;.
\end{eqnarray} 
At present the functions $G^{SE}$ have been determined numerically
at $Z=1$ and $Z=2$ \cite{17bcs}, \index{Radiative corrections!01@one-loop!self-energy}
\begin{equation}
G^{\rm SE}_{1}(Z=1) = 16.079(15)
\end{equation}
and
\begin{equation}
 G^{\rm SE}_{1}(Z=2) = 15.29(9)
\end{equation}
while $G^{VP}_{1}$ is known analytically \cite{17kis}:
\index{Radiative corrections!01@one-loop!vacuum-polarization}
\begin{equation}
G^{\rm VP}_{1} = -{8 \over 15} \,\ln(2) + {34 \over 225} + \pi (Z \alpha)
\left[ - {13 \over 24}\, \ln{ Z\alpha \over 2} + {539 \over 288}\right ] + ...
\end{equation}
To present results for the $2s$ state, we can use the results of Ref. \cite{17zwanziger} 
for $D_{21}$, which however include terms only up to order $\alpha (Z\alpha)^2 E_F$. 
After we recalculated some integrals from paper \cite{17zwanziger} the
result is
\begin{equation}
G^{\rm SE}_{2} = G^{\rm SE}_{1} + \left(-7 + {16 \over 3} \,\ln(2)\right)\, \ln(Z\alpha) -
5.221233(3) + {\cal O} \big(\pi (Z\alpha)\big)
\end{equation}
and
\begin{equation}
G^{\rm VP}_{2} = G^{\rm VP}_{1} - {7 \over 10} + {8 \over 15} \,\ln(2) + {\cal O}\big(\pi 
(Z \alpha)\big)\;.
\end{equation}

Continuing to the two-loop corrections, all terms known to date are state-independent, 
so we give only the ground state result \cite{17KNio,17ES,17ZP},
\index{Two-loop corrections}
\[
D^{(4)}_{1} = a_e^{(4)} + 0.7718(4) \,\pi (Z \alpha) - {4\over 3} \,(Z\alpha)^2\,
{\rm ln}^2(Z\alpha)\;.
\]
Non-leading terms, including single powers of ln$(Z\alpha)$ and constants,
are both state-dependent, but unknown.

When the nucleus is not point-like, the leading correction is known as the Zemach correction,
\begin{equation}
\Delta\nu_n({\rm Zemach}) = {8 E_F \over \pi n^3} \,(Z\alpha m) \,\int {dp \over p^2} \left(
G_E(p) \widetilde{G}_M(p) - 1\right)\;.
\end{equation}
Inaccuracy arisen from uncertainties in the form factors $G_E$ and $\widetilde{G}_M$, which
are both normalized to unity at zero momentum, and
from the lack of knowledge of polarization effects, is large as about 10 ppm or 4 ppm respectively, 
but those leading terms are state-independent and do not contribute into the the difference $D_{12}$.
\index{Zemach correction}

\index{Recoil corrections|(}
For atoms with nuclear structure the following result was found  \cite{17sternheim}
\begin{eqnarray}
\Delta D_{21}({\rm Rec}) = (Z\alpha)^2 \,{ m \over M}\,\Biggl\{ &-&{9 \over 8} 
+ \left[-{7 \over 32} +
{ \ln(2) \over 2}\right]\left(1-{1 \over x}\right) 
\nonumber\\
&-& \left[{145 \over 128} - { 7 \over 8} \,\ln(2) \right]\,x\Biggr\}\;,
\end{eqnarray}
where ${g M / Z m_p} = x = {(\mu / \mu_B)} \,{(M / m)}\,{ (1 / Z I)}$.
It does not depend on the nuclear structure effects such a distribution of the nuclear 
charge and magnetic moment.
Contrary, the leading recoil term for the $\Delta \nu_n$ 
(which has order $(Z \alpha)(m/M)\ln(M/m)$ \cite{17arno} 
is essentially nuclear-structure depen\-dent.
\index{Recoil corrections|)}

\section{Present status of $D_{21}$ theory}

\subsection{Old theory and recent progress}

The Breit, Zwanziger and Sternheim corrections \cite{17brei,17zwanziger,17sternheim} 
lead to a result
\begin{eqnarray} \label{17oldth}
D_{21} = E_F \,(Z\alpha)^2 
\times &\Biggl\{ &\left[{5 \over 8} + {177 \over 128}\,(Z\alpha)^2\right]
\nonumber\\
&+&{\alpha\over \pi}\,\left[
\left(-7 + {16 \over 3} \,\ln(2)\right)\, \ln(Z\alpha) -5.37(6)\right]
\nonumber\\
&+&{\alpha\over \pi}\,\left[- {7 \over 10} + {8 \over 15} \,\ln(2) \right]
\nonumber\\
&+&{m\over M}\,\biggl[
 -{9 \over 8} +  \left(-{7 \over 32} +
{ \ln(2) \over 2}\right)\left(1-{1\over x}\right) 
\nonumber\\
&-&\left({145 \over 128} 
 -{ 7 \over 8} \,\ln(2) \right)\,x\biggr]\Biggr\}\;.
\end{eqnarray} 

Some progress was achieved before we started our work. In particular, we need to mention 
two results:
\begin{itemize}
\item
Integrals, used for in Ref. \cite{17zwanziger}, were evaluated later by 
P. Mohr\footnote{Unpublished. The result is quoted accordingly to Ref. \cite{17prior}.} 
with higher accuracy and the constant was found to be -5.2212 instead of 
-5.37(6).
The theoretical 
prediction based on Eq. (\ref{17oldth}) but with a corrected value of the contstant is Table 1 as ``old theory''.
\\
\item
Some nuclear-structure- and  state-dependent corrections were found 
\cite{17kars97} for arbitrary $nS$.
\\
\end{itemize}

\subsection{Our results}

\index{Logarithmic corrections|(}
The similar difference has been under investigation also for the Lamb shift and a number of useful 
auxiliary expressions have been found for calculating the state dependent corrections to the Lamb shift 
\cite{17kars}. 

Let us mention that an improvement in the accuracy and new result on higher $n$ hfs can 
be expected with progress in 
optical measurements and we present here a progress also for higher $n$, 
defining $D_{n1}=n^3 \Delta \nu_n-\Delta \nu_1$. 
\begin{itemize}
\item
We have reproduced the logarithmic part of the self energy contribution and found for arbitrary 
$ns$ 
\index{Radiative corrections!01@one-loop!self-energy}
\begin{eqnarray}
\Delta D_{n1} &=& {\alpha \over \pi} \, (Z\alpha)^2\, E_F\,\ln(Z\alpha)\,
\left(-{8\over3}\right)
\nonumber\\
&\times&\left[2\left(-\ln(n)+{{n-1}\over n}+\psi(n)-\psi(1)\right)-{{n^2-1}\over{2n^2}}\right]\;.
\end{eqnarray}
The calculation is based on a result in Ref. \cite{17kars} for the singe logarithmic correction due to 
the one-loop self energy and one-loop vacuum polarization. 
\\
\item
We have reproduced the vacuum polarization contribution and found that for arbitrary $ns$ 
\index{Radiative corrections!01@one-loop!vacuum-polarization}
\begin{eqnarray}
\Delta D_{n1} &=& {\alpha \over \pi} \, (Z\alpha)^2\, E_F\,\left(-{4\over15}\right)
\nonumber\\
&\times&
\left[2\left(-\ln(n)+{{n-1}\over n}+\psi(n)-\psi(1)\right)-{{n^2-1}\over{2n^2}}\right]\;.
\end{eqnarray}
\\
\item
Integrals used by Zwanziger \cite{17zwanziger} have been recalculated 
and the constant was found to be -5.221233(3).
\\
\item
We also found two higher-order logarithmic corrections
\index{Two-loop corrections}
\begin{eqnarray}
\Delta D_{n1} &=& {\alpha^2 \over \pi^2} \, (Z\alpha)^2\, 
E_F\,\ln(Z\alpha)\,\left(-{4\over3}\right)\nonumber\\
&\times&
\left[2\left(-\ln(n)+{{n-1}\over n}+\psi(n)-\psi(1)\right)-{{n^2-1}\over{2n^2}}\right]
\end{eqnarray}
and
\begin{eqnarray}
\Delta D_{n1} &=& {\alpha \over \pi} {m\over M}\, (Z\alpha)^2\, 
E_F\,\ln(Z\alpha)\,\left({16\over3}\right)\nonumber\\
&\times&
\left[2\left(-\ln(n)+{{n-1}\over n}-\psi(n)+\psi(1)\right)-{{n^2-1}\over{2n^2}}\right]\;.
\end{eqnarray}
\index{Logarithmic corrections|)}\\
\item
We found two higher-order non-logarithmic corrections
\begin{eqnarray}
\Delta D_{n1}^{SE} &=& \alpha \, (Z\alpha)^3\, E_F \,
\Bigg\{\left[{139\over 16}-4\ln(2)\right]
\nonumber\\
&\times&
\left[
-\ln(n)+
{{n-1}\over n}+
\psi(n)-\psi(1)
\right]
\nonumber\\
&+&\left[\ln(2)-{13\over 4}\right]\nonumber\\
&\times&
\left[\psi(n+1)-\psi(2)-\ln(n)-\frac{(n-1)(n+9)}{4n^2}\right]
\Bigg\}
\end{eqnarray}
and
\begin{eqnarray}
\Delta D_{n1}^{\rm VP} &=& \alpha \, (Z\alpha)^3\, E_F \times
\Bigg\{\left[{5\over 24}\right]
\left[-\ln(n)+{{n-1}\over n}+\psi(n)-\psi(1)\right] 
\nonumber\\
&+&\left[{3\over 4} \right]\times
\left[\psi(n+1)-\psi(2)-\ln(n)-\frac{(n-1)(n+9)}{4n^2}\right]
\Bigg\}
 \;.
\end{eqnarray}
\index{Radiative corrections!01@one-loop!vacuum-polarization}\\
\item
We have also found a term proportional to the magnetic radius.
To the best of our knowledge
that is the first contribution, which is proportional to
the magnetic radius and on the level of the experimental accuracy.
The complete nuclear-structure correction is
\begin{eqnarray}
\Delta D_{n1}^{\rm Nucl}&=&-(Z\alpha)^2\,
\left[\psi(n+1)-\psi(2)-\ln(n)-\frac{(n-1)(n+9)}{4n^2}\right]\nonumber\\
&\times&\Delta \nu_{1}({\rm Zemach + polarizability})
+{4\over 3}(Z\alpha)^2\,\biggl[
\psi(n)-\psi(1) -\ln(n) 
\nonumber\\
&+&\frac{n-1}{n}-\left(\frac{R_M}{R_E}\right)^2\frac{n^2-1}{4n^2}\biggr]\,
\Big(m R_E\Big)^2\,E_F\;.\nonumber\\
\end{eqnarray}
\end{itemize}
\index{Nuclear structure!electric charge distribution}
\index{Nuclear structure!magnetic moment distribution}
\index{Proton structure!magnetic moment distribution}
\index{Proton structure!electric charge distribution}
\index{Zemach correction}
\index{Nuclear structure!nuclear polarizability}
\index{Proton structure!polarizability}

\section{Present status}

To calculate the corrections presented in the previous sections, we have used an effective non-relativistic 
theory. In particular we have studied two kinds of terms. One is a result of the second order perturbation 
theory with two $\delta({\bf r})$-like potentials, evaluated in the leading non-relativistic 
approximation, while the other is due to a more accurate calculation of a single delta-like potential. 
Both kinds contribute into the state-independent leading logarithmic corrections 
($\alpha^2(Z\alpha)^2\ln^2(Z\alpha) $, $\alpha(Z\alpha)^2(m/M)\ln^2(Z\alpha)$, 
and $\alpha(Z\alpha)^3\ln(Z\alpha) $) 
and to next-to-leading state-dependent terms 
($\alpha^2(Z\alpha)^2\ln(Z\alpha) $, $\alpha(Z\alpha)^2(m/M)\ln(Z\alpha) $, 
and $\alpha(Z\alpha)^3$).
The crucial question is if we found all corrections in these orders. 
Rederiving a leading term in order $\alpha(Z\alpha)^2\ln(Z\alpha)$ within our technics,
we can easily incorporate 
the anomalous magnetic moment of the electron in the calculation and restore the nuclear mass dependence.
Since we reproduce the well-known result for the $\alpha(Z\alpha)^2\ln(Z\alpha)$ 
term, we consider that as a confirmation of two other logarithmic corrections found by us. In the case of 
$\alpha(Z\alpha)^3$ it might be a contribution of an effective operator, proportional to 
$(\Delta/m)\delta({\bf r})$. That can give no logarithmic corrections, but leads to a state-dependent 
constant. We are now studying this possibility.

Summarizing all corrections, the final QED result is found to be:
\begin{eqnarray}
D_{21}^{\rm QED} &=& E_F \,(Z\alpha)^2 
\times \Biggl\{ \left[{5 \over 8} + {177 \over 128}\,(Z\alpha)^2\right]
\nonumber\\
&+& {\alpha\over \pi} \, \left[
\left(-7 + {16 \over 3} \,\ln(2)\right)\, \ln(Z\alpha) -5.221233(3)\right]
\nonumber\\
&+&{\alpha\over \pi}\,\left[- {7 \over 10} + {8 \over 15} \,\ln(2) \right]
\nonumber\\
&+& {m\over M}\,\biggl[
 -{9 \over 8} +  \left[-{7 \over 32} +
{ \ln(2) \over 2}\right]\left(1-{1\over x}\right) 
- \left[{145 \over 128} - { 7 \over 8} \,\ln(2) \right]\,x\biggr]\bigg\}
\nonumber\\
&+&{\alpha^2\over 2\pi^2} \,\left(-7 + {16 \over 3} \,\ln(2)\right)\, 
\ln(Z\alpha)
\nonumber\\
&-&{\alpha\over \pi}{2m\over M} \, \,\left(-7 + {16 \over 3} \,\ln(2)\right)\, 
\ln(Z\alpha)
\nonumber\\ 
&+& \alpha(Z\alpha)
\Bigg\{\left[{139\over 16 }-4\ln(2)+{5\over 24}\right]
\left[ {3\over 2}-\ln(2) \right] 
\nonumber\\
&+&\left[ {13\over 4} -\ln(2)-{3\over 4} \right]
\left[ \ln(2)+{3\over 16}\right] \Biggr\}
\;.
\end{eqnarray}

Numerical results (in kHz) for hydrogen and deuterium atoms and the helium-3 ion are collected in Table 2.
One can see that the new corrections essentially shift the theoretical predictions. A comparison of 
the QED predictions (in kHz) against the experiments is summarized in Table 1.
\index{Deuterium atom}
We take the values 
of the fundamental constants (like e. g. the fine structure constant $\alpha$) 
from 
the recent adjustment (see Ref. \cite{17mohr}).
\begin{table}
\caption{QED contributions to the $D_{21}$ in hydrogen, deuterium and helium-3 ion. 
The results are presented in kHz}
\label{17Tab1}
\begin{center}
\def\arraystretch{1.4}
\setlength\tabcolsep{5pt}
\begin{tabular}{lccc}
\hline
Contribution                        &  H      & D        & $^3$He$^+$  \\
\hline
\phantom{+} $(Z\alpha)^2 E_F$        & 47.2275 &  10.8835 & 1\,152.9723 \\
+ $\alpha(Z\alpha)^2 E_F $ (SE)      &  1.9360 &   0.4461 & 37.4412     \\
+ $\alpha(Z\alpha)^2 E_F $ (VP)      & -0.0580 &  -0.0134 & -1.4148     \\
+ $(Z\alpha)^2{m\over M} E_F $       & -0.1629 &  -0.0094 & 0.7966      \\
\hline
+ $\alpha^2(Z\alpha)^2 E_F $         &  0.0033(16) &   0.0008(4) &    0.070(35) \\
+ $\alpha(Z\alpha)^2{m\over M} E_F $ & -0.0031(15) &  -0.0004(2) &    -0.022(11) \\
+ $\alpha(Z\alpha)^3 E_F $ (SE)      &  0.0282 &   0.0065 &    1.3794 \\
+ $\alpha(Z\alpha)^3 E_F $ (VP)      & -0.0019 &  -0.0005 &    -0.0967 \\
\hline
\end{tabular}
\end{center}
\end{table}

An important point is that the difference is sensitive to 4th order corrections and so is
competitive with the muonium {\em hfs} as a test of the QED. 
\index{Bound state QED!precision tests}\index{Muonium!hyperfine structure}%
The difference between the QED part of 
the theory and the experiment is an indication of higher-order 
corrections due to the QED and the nuclear structure, which have to be studied in 
detail.
In particular, we have to mention that while we expect that we have a complete result 
on logarithmic corrections and on the vacuum-polarization part of the $\alpha(Z\alpha)^3$ term 
we anticipate more contributions in the order $\alpha(Z\alpha)^3$ due to the self-energy. 
A complete study of this term offers a possibility to determine the magnetic radius of the proton, 
deuteron and helion-3. \index{Deuterium atom} 

\index{State-dependent corrections|)}
\index{Hydrogen atom!hyperfine structure|)}
\index{Helium ion!hyperfine structure in $^3$He$^+$|)}

\section*{Acknowledgments}

The author would like to thank Mike Prior, Dan Kleppner and especially Eric Hessels
for stimulating discussions. An early part of this work was done during my short but 
fruitful stay at University of Notre Dame and I am very grateful to Jonathan Sapirstein 
for his hospitality, stimulating discussions and participation in the early stage of this project. 
The work was supported in part by RFBR grant 00-02-16718, NATO grant CRG 960003 and by
Russian State Program ``Fundamental Metrology''.

\label{c_kar1_}

\end{document}